\documentclass{svproc}

\usepackage{graphicx}
\usepackage{multirow}
\usepackage{amsmath,amssymb,amsfonts}
\usepackage{mathrsfs}
\usepackage{xcolor}
\usepackage{textcomp}
\usepackage{manyfoot}
\usepackage{booktabs}
\usepackage{algorithm}
\usepackage{algorithmicx}
\usepackage{algpseudocode}
\usepackage{tikz}
\usetikzlibrary{arrows.meta,positioning}
\usepackage{url}

\newcommand{\worstof}{\mathrm{worst\_of}}
\newcommand{\bestof}{\mathrm{best\_of}}

\definecolor{cascgreen}{HTML}{22C55E}
\definecolor{cascred}{HTML}{EF4444}
\definecolor{cascring}{HTML}{FBBF24}
\definecolor{casclbl}{HTML}{52525B}
\newcommand{\legdia}[1]{\tikz[baseline=-0.7ex]{\fill[#1] (-2.4mm,0)--(0,2.4mm)--(2.4mm,0)--(0,-2.4mm)--cycle;}}
\newcommand{\legoct}[1]{\tikz[baseline=-0.7ex]{\fill[#1] (22.5:2.6mm) \foreach \a in {67.5,112.5,157.5,202.5,247.5,292.5,337.5}{--(\a:2.6mm)}--cycle;}}
\newcommand{\legcir}[1]{\tikz[baseline=-0.7ex]{\fill[#1] (0,0) circle (2.2mm);}}
\newcommand{\legcrack}{\tikz[baseline=-0.7ex]{\fill[cascred] (0,0) circle (2.2mm);
  \draw[white,line width=.7pt] (-0.6mm,2.1mm)--(0.4mm,0.4mm)--(-0.5mm,-0.3mm)--(0.5mm,-2.1mm);}}
\newcommand{\legring}{\tikz[baseline=-0.7ex]{\draw[cascring,line width=.9pt] (0,0) circle (2.6mm);\fill[cascgreen] (0,0) circle (1.7mm);}}
\newcommand{\legicon}[1]{\raisebox{-0.55ex}{\includegraphics[height=2.5mm]{#1}}}
\usepackage[hidelinks]{hyperref}
\begin{document}
\mainmatter 

\title{Composable, Knowledge-Driven Disservice Propagation for Complex Interdependent Services with a Flow-Allocation Module Validated Against Hydraulic Simulation}

\titlerunning{Composable Knowledge-Driven Disservice Propagation}

\author{Cristian Curaba \and Stefano Grimaz}
\authorrunning{C. Curaba and S. Grimaz}
\institute{University of Udine, Udine, Italy,\\
\email{cristian.curaba@uniud.it}}

\maketitle

\begin{abstract}
Modeling cascading failure across interdependent critical infrastructure requires either
a composable physical simulator per domain or a topological abstraction that discards
quantity awareness. Simulators are costly, data-hungry, and require detailed calibration per domain; topological
abstractions stay data-parsimonious while missing quantity-based behaviors that govern real dynamics. 
We present, instead, a knowledge-driven propagation architecture in which heterogeneous
service dependencies are identified by categories: a general logical propagation resolves
multi-dependency and redundancy structures purely from how nodes are categorized, while
quantity-aware categories plug in dedicated mechanisms---here, a max-min fair-share flow
module for water distribution. Both share one discrete functionality scale and a
monotonic update algorithm guaranteeing convergence in finitely many rounds. 
We assess the flow module by comparing its results with those we get from preexisting
pressure-driven simulation with WNTR (Water Network Tool for Resilience) on eight realistic networks across four failure families (clustered damage, capacity-targeted attack, source failure,
tank loss). Under capacity-targeted attack and source failure a topological-reachability baseline misses
46\% of the truly critical junctions, where the flow module misses 10\% and 8\% at an average critical-class precision of
0.73.\footnote{Supplementary materials, code and results are available at:
\url{https://github.com/cascade-platform-org/CASCADE}. The platform itself runs at \url{https://app.cascade-platform.org/}.}
\end{abstract}

\section{Introduction}\label{sec:intro}

Hazards frequently trigger cascading failures across interdependent
critical infrastructure, amplifying initial damage well beyond the
footprint of the initiating event---from Hurricane Maria's long-lasting
multi-system service crisis in Puerto Rico~\cite{hurricane_maria} to the
April 2025 Iberian blackout's national-scale propagation into transport
and digital services~\cite{panel2025grid}. The UN Office for Disaster Risk Reduction (UNDRR) resilient-infrastructure handbook
calls for tools that capture ``the interdependent nature of infrastructure and services
[\ldots] and the cascading impacts''~\cite{UNDRR2023Handbook}.

Existing computational approaches sit at two extremes. Physical simulators reproduce one
domain's dynamics with high fidelity, at the price of extensive domain-specific
calibration that does not apply to another domain's physics. Topological and
network-of-networks models generalize across domains on structural information alone, while
discarding the quantity-aware dynamics---partial supply, congestion, priority-driven
allocation---that decide whether a service is actually delivered.

Models need both readings, for different dependencies. Whether a district still
has water is a question of \emph{how much}: a limited supply divides across contended
mains. Whether its pumping station still runs is a question of \emph{whether}: it needs
electricity and telemetry which generally are either available or not. No framework we know of
resolves both kinds inside a single architecture that a domain expert extends by
declaring just a few node attributes, rather than by writing/integrating a new solver for each kind of
dependency.

We present a category-typed architecture for propagating service
disruptions through directed infrastructure networks. Nodes carry a
discrete functionality level, while categories specify whether a
dependency is logical or quantity-aware. Logical propagation models
requirements and redundancy using worst-of and best-of aggregation.
For consumable resources, a plug-in flow module allocates finite supply
among demand nodes using priority-tiered max-min fairness. Both
mechanisms share the same functionality scale and monotonic update rule,
which guarantees finite convergence. We instantiate the flow module for
water distribution, automatically importing EPANET (the standard open hydraulic solver) networks and
comparing the resulting functionality estimates with pressure-driven
WNTR simulations. The module is intended as a light-data approximation,
not as a replacement for detailed hydraulic analysis.

\paragraph{Contributions.} (1) A composable, category-based propagation
architecture in which quantity-aware and purely logical interdependencies
share a single convergent update rule (Section~\ref{sec:architecture}). 
(2) A pipeline to automatically label node and edge attributes exploting information gained from hydraulic
solutions (Section~\ref{sec:importer}). (3) A validation of that module
against pressure-driven WNTR simulation used as the hydraulic reference on eight networks (Section~\ref{sec:validation}).
\section{Related Work}\label{sec:related}

Foundational work on interdependent infrastructure established the
classification of interdependencies into physical, cyber, geographic, and
logical dimensions~\cite{rinaldi_identifying_2002}; the
modeling-and-simulation literature that followed spans empirical,
agent-based, economic, and network-based
approaches~\cite{ouyang_review_2014}. Subsequent network-of-networks
research showed that even weak inter-system coupling
can trigger abrupt, catastrophic collapse rather than gradual degradation
\cite{buldyrev_catastrophic_2010,gao_networks_2011}.
These studies characterize \emph{when} cascades occur but stop short of a
computational mechanism a practitioner can run on an arbitrary network; our
logical propagation operationalizes exactly that. In
multilayer-network terms~\cite{kivela_multilayer_2014,boccaletti_structure_2014,curaba2026seriesparalleldependencylinks},
our categories play the role of layers to model the disservice propagation across them.

Networks are far more vulnerable to targeted removal of their most important elements than
to random loss~\cite{albert_error_2000} which is the regime that matters for
resilience assessment, and the harder one to predict. Three of our four failure
families are deterministic attacks of that kind, aimed at main pipes and pumps and at storage (Section~\ref{sec:validation}).

Load-redistribution models~\cite{motter_cascade-based_2002,crucitti_model_2004} treat a
different mechanism: losing one element reroutes its load onto others, and where
that load exceeds a survivor's capacity the survivor fails in turn---modelled
for water distribution~\cite{shuang_performance_2014} and for
interdependent water--power systems~\cite{li_cascading_2025}. Our module
allocates capacitated flow against demand declared at explicit sources, so a
shortfall surfaces as unmet delivered quantity (Section~\ref{sec:discussion}).
Network-flow formulations of interdependent infrastructure are well established for service
restoration~\cite{lee_restoration_2007}, and broader resilience
reviews~\cite{de_marco_quantitative_2025,hines_topological_2010} converge on the
gap this paper addresses: topological metrics are data-parsimonious but
function-blind, while performance-based models capture function at the cost of
per-domain simulation machinery with no practical way of doing comprehensive evaluations.

For water-network tooling, WNTR~\cite{klise_software_2017} implements the
pressure-driven demand (PDD) formulation~\cite{wagner_water_1988} we use
as ground truth (Section~\ref{sec:validation}); our importer's optional
node-budget reduction uses its skeletonization by diameter
threshold~\cite{saldarriaga_water_2012}. EPANET~\cite{rossman2000epanet}
is the solver both are built on. Graph-theoretic robustness measures for water
networks~\cite{yazdani_complex_2011,herrera_graph_2016} are the closest
precedent for our reachability baseline: they judge a network from its topology
alone, which costs nothing to compute but says nothing about how much water is
actually delivered. How well they predict has been shown to depend on where the
sources sit~\cite{meng_topological_2018}. Machine-learned
surrogates~\cite{garzon_machine_2022} occupy the same niche from the
opposite direction---capturing hydraulics but needing per-network
training data, where our flow module is built from
one import of the network file itself. Our flow module's
allocation builds on classical max-min fairness
principle~\cite{nace_tutorial_2006} standard in network resource
allocation, implemented via NetworkX~\cite{hagberg_exploring_2008}.

\section{The Propagation Architecture}\label{sec:architecture}

We introduce the architecture through one small example
(Section~\ref{sec:arch-example}), then make each of its parts precise: the elements a
model is built from and how their categories are typed
(Section~\ref{sec:arch-model}), the logical propagation (Section~\ref{sec:arch-logical}) and the flow module
(Section~\ref{sec:arch-flow}), and finally the shared update rule that composes them
and guarantees convergence (Section~\ref{sec:arch-pipeline}).

\subsection{Running Example: An Office}\label{sec:arch-example}

\begin{figure}[!t]
\centering
\begin{minipage}[b]{0.44\textwidth}\centering
{\scriptsize (a) after the blackout event}\\[1pt]\includegraphics[width=\linewidth]{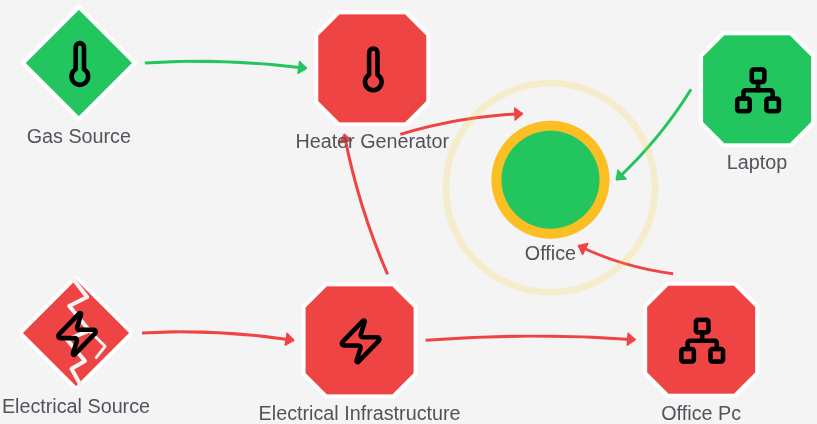}
\end{minipage}\hfill
\begin{minipage}[b]{0.44\textwidth}\centering
{\scriptsize (b) under a shared digital fault}\\[1pt]\includegraphics[width=\linewidth]{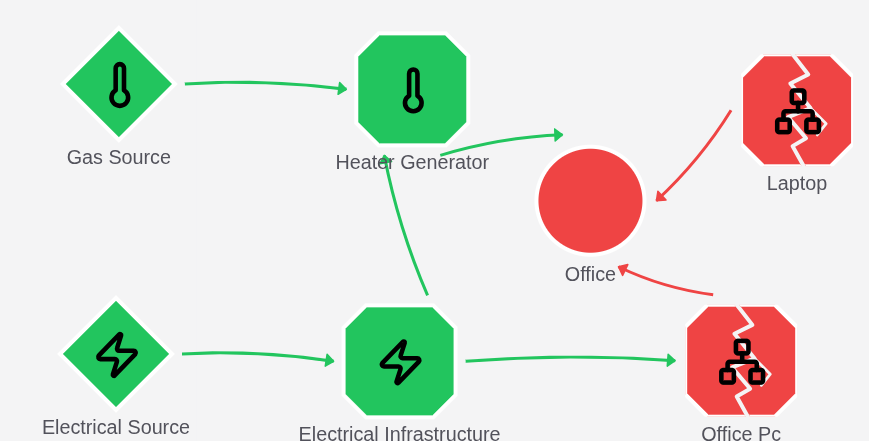}
\end{minipage}

\vspace{3pt}
{\centering\scriptsize
\legdia{casclbl!55}\,Source \quad
\legoct{casclbl!55}\,Infrastructure \quad
\legcir{casclbl!55}\,Service\\[2pt]
\legcir{cascgreen}\,operational \quad
\legcir{cascred}\,critical \quad
\legcrack\,struck by the event \quad
\legring\,backup defers the drop\\[2pt]
\legicon{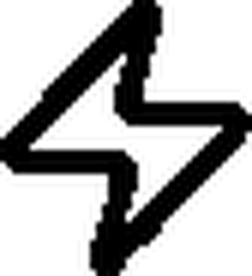}\,\emph{electric} \quad
\legicon{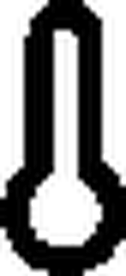}\,\emph{heat} \quad
\legicon{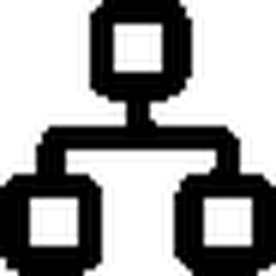}\,\emph{digital}\par}
\vspace{3pt}

\caption{The running example as direct screenshots, both starting from an all-operational initial condition.
 In~(a) the cracked Electrical Source zeroes the electric supply, so the Office~PC
shuts down and the Heater Generator drops by \emph{worst-of} across categories
(gas alone cannot run it). The Office itself stays functional, for two distinct
reasons: \emph{best-of} keeps its digital dependency alive through the surviving
Laptop, and its heat backup defers the drop for as long as the room stays warm
(yellow ring). In~(b) a shared software fault instead takes \emph{both} digital
devices critical: \emph{best-of} now has no surviving digital parent, and the
Office goes critical}\label{fig:example}
\end{figure}

Figure~\ref{fig:example} shows an office with common dependencies: the served office node depending on two \textsc{SourceToDemands} categories---\emph{electric} (an \textsc{Electrical Source} feeding an
\textsc{Electrical Infrastructure}) and \emph{heat} (a \textsc{Gas Source} feeding
a \textsc{Heater Generator})---and one \textsc{Requisite} category, \textsc{digital},
through two independent devices, a \textsc{Laptop} and an \textsc{Office PC}.

\emph{Redundancy vs Requirements.} Losing either digital device alone leaves the
office's digital dependency intact, one surviving device being enough; the
logical propagation encodes this by aggregating the functionality of
parent nodes (predecessors, joined by an edge oriented towards the node) within a
category via \emph{best-of}
(Section~\ref{sec:arch-logical}). The office, however, requires electricity,
heat, \emph{and} digital service together, which is \emph{worst-of} aggregation
across categories.

\emph{Quantity.} What the logical propagation leaves out is amount: whether a
source can cover every downstream demand concurrently. That is what
Section~\ref{sec:arch-flow}'s flow module supplies, and why
\textsc{SourceToDemands} categories route through it as well. The example is
deliberately minimal---one source per resource, no alternative routing---so
the flow module adds no behaviour here. Amount becomes decisive on a
\emph{capacitated} network with \emph{several} source-to-demand paths, the regime
the water networks of Section~\ref{sec:validation} exercise.

\subsection{Elements, Functionality, and Category Typing}\label{sec:arch-model}

A modeled network is a directed graph, with nodes and edges each carrying an integer \emph{functionality} on a discrete scale
$1,\ldots,N$ ($1$ = critical, $N$ = full operation); small discrete
state scales are standard in disaster-loss modeling (e.g.\ Hazus damage
states~\cite{kircher_hazus_2006}), and $N$ is a modeling choice (defaulting to 3).
Every node is attributed with a set of \emph{categories}---named services such as
\emph{water}, \emph{electric}, or \emph{digital}---which represent the service it provides (generally, just one)
 and every category is \emph{typed}, which selects how that dependency is propagated.

Two algorithms do the proposal functionality updates. The \emph{logical} propagation runs
for every node over all of its incoming edges as defined in
Section~\ref{sec:arch-logical}. Nodes without positive demands and labelled with a category typed \textsc{Requisite} is resolved by only
the logical propagation. While nodes with positive demands or with a category typed \textsc{SourceToDemands} are considered in the flow module,
which solves one capacitated allocation over the considered subgraph (Section~\ref{sec:arch-flow}). A node holding a proposal from both then
takes the worse of the two (Section~\ref{sec:arch-pipeline}).

Neither algorithm sets a functionality directly: each \emph{proposes} one, and per-node
attributes may adjust that proposal before it is committed. A category's
\emph{dependency level} attenuates how hard a drop in that category hits the node, and
a \emph{backup} defers the drop by a specified amount of time (as \textsc{Office} in
Figure~\ref{fig:example}). These guards and the commit that follows them are the subject
of Section~\ref{sec:arch-pipeline}.

  \subsection{Logical Propagation}\label{sec:arch-logical}

  For a target node $v$ with incoming edges $E_{\mathrm{in}}(v)$, let $(u\!\to\!v)\in E_{\mathrm{in}}(v)$ be one such
  edge, $u$ its parent node, $s(u)\in\{1,\ldots,N\}$ that parent's functionality and
  $s(u\!\to\!v)$ the edge's own. The logical propagation adopts two primitive combinators
  on the totally ordered scale: $\worstof(x,y)=\min(x,y)$, modeling \emph{requirement}
  (every input must hold for the output to hold), and $\bestof(x,y)=\max(x,y)$, modeling
  \emph{redundancy} (one healthy input in the same category suffices). The logical propagation proposal
  is the result of the following three steps:
  \begin{enumerate}
    \item \textbf{Deliverable functionality}---a degraded parent and a
      degraded connecting edge both limit what can reach $v$:
      $L(u\!\to\!v)=\worstof\big(s(u),\,s(u\!\to\!v)\big)$.
    \item \textbf{Intracategorical aggregation}---within one category $g$,
      one healthy supplier is enough:
      $C_g(v)=\bestof\big\{L(u\!\to\!v):(u\!\to\!v)\in E_{\mathrm{in}}^g(v)\big\}$, 
      where $E_{\mathrm{in}}^g(v)\subseteq E_{\mathrm{in}}(v)$ are the incoming edges whose parent supplies category $g$.
    \item \textbf{Intercategorical aggregation}---$v$ requires \emph{every}
      category it depends on, so categories combine by requirement:
      $s^{\mathrm{logic}}(v)=\worstof\big\{C_g(v):g\in G(v)\big\}$, with $G(v)$ the
      categories represented among $v$'s parents.
  \end{enumerate}
  We notice that each of the above is a deterministic function of the parent nodes'
  and edges' functionalities, making the proposal locally computable.
  The two combinators are the many-valued generalization of classical fault-tree
  \textsc{and}/\textsc{or} gates~\cite{vesely1981fault}: a node's proposal is a
  deterministic function of its parents' functionality with no free parameter, computable
  by inspection on the running example.

  \subsection{Flow Propagation Module}\label{sec:arch-flow}

  A category typed \textsc{SourceToDemands} represents a service based on a resource that is sourced, carried and
  consumed in continuous amounts. The resource behaviour can differ depending on the domain: it may be governed by physical laws (like in a water distribution system),
  or by contractual agreements in a supply chain.
  This section evaluates how closely the hydraulically informed flow module reproduces pressure-driven WNTR 
  results. The importer derives the module's graph structure and capacities from EPANET 
  data and auxiliary hydraulic solutions; the resulting abstraction is then compared with PDD-WNTR 
  simulations of the same networks under the tested failures.

The algorithm's goal is to determine, for every consuming node, the largest share of 
its declared demand the network's scarce supply can deliver once higher-priority 
consumers are served in full, dividing any remaining shortage among equal-priority 
consumers as evenly as the network's capacities permit and turning the result into 
a per-node functionality estimate.
  For a typed \textsc{SourceToDemands} category, the module extracts the subgraph 
  (induced by all nodes with positive demands or labelled with that category)
  and solves one priority-aware capacitated flow problem on it. Within that subgraph,
  \emph{source} nodes supply the resource at a rate scaled linearly by their own
  functionality, \emph{infrastructure} nodes relay it under a throughput capacity, and
  \emph{service} nodes only consume it, while edges carry a capacity. The module solves this on an auxiliary
  directed graph. Every node is split into an \emph{in} and an \emph{out} vertex joined by
  an edge carrying that node's throughput, so a node's own capacity is enforced like any
  other edge capacity, and the original edges run from one node's \emph{out} vertex to the next
  node's \emph{in} vertex. A super-source feeds each supplying node at its effective
  supply, and each demanding node drains into a super-sink at its declared demand. It allocates flow across this
  auxiliary graph by \emph{priority-tiered max-min fairness}~\cite{nace_tutorial_2006}: higher tiers
  served fully first, equal-priority consumers sharing any shortage as evenly as
  the network physically allows. For each consuming node, the delivered
  fraction $r$ of declared demand quantizes to a proposed functionality
  $s^{\mathrm{cap}}(v)=\max(1,\lceil r N\rceil)$. 
  Priority is a \emph{modelling} input---``serve $X$ before $Y$'' from expert knowledge of how the system is actually
  operated.

  \subsection{Shared Update Rule and Convergence}\label{sec:arch-pipeline}

  Every round applies the same three steps to every node---\emph{propose},
  \emph{guard}, \emph{commit}---whichever propagation is involved.

  \emph{Propose.} Section~\ref{sec:arch-logical} yields $s^{\mathrm{logic}}(v)$ for every
  node, and Section~\ref{sec:arch-flow} yields $s^{\mathrm{cap}}(v)$ for each node
  with a \textsc{SourceToDemands} category. Both live on the same $1,\ldots,N$
  scale, so a node holding both aggregates them by necessity,
  $s^{\mathrm{prop}}(v)=\worstof\big(s^{\mathrm{logic}}(v),\,s^{\mathrm{cap}}(v)\big)$,
  while a node holding only one takes it unchanged.

  \emph{Guard.} Two per-category attributes may then soften that proposal, carrying
  $s^{\mathrm{prop}}(v)$ into the guarded value $s^{\mathrm{guarded}}(v)$ the commit
  consumes. A guard only modulates a drop the mechanisms have already proposed,
  and a node that declares nothing gets the worst case (representing full dependency and no
  reserve). The \emph{dependency level} $d\in\{1,\ldots,N\}$ states how hard the node
  leans on a category, attenuating a proposed level $\ell$ to $\min(N,\,\ell+(N-d))$: at
  $d=N$ the drop passes in full, at $d=1$ it is neutralized, and values between reduce it
  linearly. A \emph{backup} then defers whatever survives attenuation: a node with a declared reserve
  in the dropping category holds its functionality and starts a countdown of that
  reserve's duration, so the drop lands only when the reserve expires. A backup covers its
  own category alone, so a simultaneous drop through an unbacked category still commits
  this round.

  \emph{Commit.} The guarded proposal is committed monotonically:
  \[
    s^{\mathrm{new}}(v) = \worstof\big(s^{\mathrm{old}}(v),\, s^{\mathrm{guarded}}(v)\big).
  \]
  Functionality can therefore only hold or worsen, never improve. This single inequality
  is what guarantees convergence.
  Every round lowers at least one element by at least one level until convergence is reached, and
  the total descent available across the model is $(N-1)(|V|+|E|)$, so the iteration
  terminates after at most $(N-1)(|V|+|E|)+1$ rounds whatever the topology and attributes---in
  practice far sooner, since a round updates every element in parallel.

  \section{Building the Flow Module from Data: The EPANET Importer}\label{sec:importer}

  Water utilities exchange networks as EPANET \texttt{.inp} files, which carry the
  physical detail of the aqueduct infrastructure. We wrote a pipeline that reads
  such a file and produces a categorized node/edge model with no manual step. Four
  decisions carry the module's fidelity; Supp.~Mat.~S1 specifies every stage in full.

  \emph{What becomes a source, and how much it supplies.} An EPANET file distinguishes two kinds
  of water body: a \emph{reservoir} is large enough to treat as inexhaustible while a \emph{tank} is a
  finite store, capped at the rate it is seen delivering while the network is intact.
  Consuming junctions become nodes with reported demand. Pumps and valves
  become inline \textsc{Requisite}-category nodes, so a stopped pump gates everything downstream
  through the logical propagation rather than through device-specific code. Pipes become
  capacitated directed edges, and demand is read at the busiest hour of the day.

  \emph{Which way the edges point.} A file records one direction per pipe, but that is
  only the direction in the operating state the file describes; elsewhere in the day, or
  after a failure, water can run the other way. We recover direction from behaviour
  instead through various \emph{hydraulic} solutions---raising demand, and
  closing one link at a time---and we read each pipe's direction from the sign of its
  simulated flow. A pipe seen flowing both ways
  becomes two opposed directed edges, as does one whose only observed flow is too slow to
  trust: committing to a direction that is solver noise can silently cut off
  everything downstream.

  \emph{How much each pipe carries.} A file gives a pipe's diameter but not its capacity.
  We apply one design velocity to every pipe, $v_{\text{design}}=2.5$\,m/s, the standard
  figure for water mains, so a pipe of diameter $d$ carries
  $\tfrac{\pi}{4}d^2 v_{\text{design}}$. Finer per-pipe assumptions do not improve fidelity (Supp.~Mat.~S5).

  \emph{Who gets water first is left unset.} Deriving that ordering from the network's own
  behaviour under failure gave no fidelity gain (Supp.~Mat.~S7), so the importer assigns
  none and every number in Section~\ref{sec:validation} uses a flat allocation, equal
  consumers sharing a shortage evenly. Priority remains central to \emph{authoring}. An
  expert who knows how a utility distributes water states that knowledge in one attribute
  per node. Where valve control makes the ordering a real operational choice, the same
  attribute serves a second purpose: comparing competing policies under shortage
  (Section~\ref{sec:discussion}).

  \section{Module Validation Against Hydraulic Simulation}\label{sec:validation}

  This section validates the flow module's fidelity to real hydraulics: the importer builds
  the module from EPANET files, and its output is compared to a pressure-driven WNTR solutions
  on the same network.

  \emph{Validation Protocol.} For each of eight networks (three textbook WNTR examples
  (Net1, Net2, Net3), three real, anonymized Italian aqueducts, and two the public
  Modena and C-Town \cite{kentucky_water_resources_research_institute_university_of_kentucky_water_nodate})
  we generate structural-failure situations,
  apply the identical intervention to a pressure-driven (PDD) WNTR solve and to
  the imported model, and compare per-junction functionality. Both sides take
  demand at the coincident system peak (Section~\ref{sec:importer}). We report three
  quantities. The \emph{Functionality Match Score} (FMS) measures agreement on
  \emph{how much} service each junction retains. Fix one situation, and let $J$ be its
  demand-bearing junctions, each $j \in J$ carrying declared demand $d_j$ and $\ell^{\mathrm{WNTR}}_j$, $\ell^{\mathrm{module}}_j$
  respectively the functionality levels (linearly mapped fraction of the demand satisfied) of the WNTR solve and the module,
  both in $\{1,\ldots,N\}$. Then
  \[
    \mathrm{FMS} \;=\; 1 \;-\; \sum_{j \in J} w_j\,
    \frac{\bigl|\ell^{\mathrm{WNTR}}_j - \ell^{\mathrm{module}}_j\bigr|}{N-1},
    \qquad w_j = \frac{d_j}{\sum_{k \in J} d_k}.
  \]
  Each junction contributes its level disagreement, divided by $N-1$, the largest the functionality scale
  allows, and weighted by $w_j$, its share of the situation's demand. FMS is $1$ on exact
  agreement and $0$ when the two sides differ by the full scale on every unit of demand.

  We evaluate at $N=3$ (critical / degraded / operational), the coarsest scale that still
  separates partial from total service loss. Agreement moves by less than 0.005 across functionality scale sizes and service-pressure conventions (Supp. Mat. S6).

  The second and third quantities look only at the \emph{critical} class, level $1$,
  against everything above it. For one situation let $W$ be the junctions the WNTR
  solve reports critical and $M$ those the module flags. Then
  \[
    \mathrm{precision} = \frac{|M \cap W|}{|M|},
    \qquad
    \mathrm{recall} = \frac{|M \cap W|}{|W|}.
  \]
  Precision and recall are the standard classification measures applied to the critical class, and FMS is the
  demand-weighted level disagreement defined above. Recall asks what share of the genuinely critical junctions were found by the module; precision asks
  what share of the module's critical flags were real. Both are computed \emph{per situation} and then
  averaged over situations. A
  fifth of our situations are ones in which the network's redundancy absorbs the
  disruption entirely, leaving $W$ empty; a model that correctly flags nothing scores
  recall $1$ there. Symmetrically a model that flags nothing at all,
  $M$ empty, scores precision $1$.

  For conservative resilience analysis recall is the
  metric that matters most. We score two baselines on the identical junction set: a
  \emph{null model} predicting every junction fully operational, and a \emph{reachability
  baseline}---the field's standard connectivity
  abstraction~\cite{hines_topological_2010}---predicting a junction critical iff no
  undirected path of in-service links connects it to a surviving source. 

  We test four failure families, each a set of simultaneous element closures (one, two or
  three at a time, capped $\sim\!60$ situations per family): \emph{cluster}, breaks within
  two topological hops of a random epicentre (local hazard);
  \emph{targeted}, among the largest-diameter mains and pumps (top fifth); \emph{source},
  closing a reservoir's outlet; and \emph{tank}, closing every link on a tank. A closed
  link carries zero flow. Table~\ref{tab:familydetect} pools the four
  families over networks and seeds; Supp.~Mat.~S4 breaks them down per network and
  Supp.~Mat.~S2 gives the complete protocol.

  \emph{FMS: the baselines nearly saturate it.} Because most sampled
  situations perturb only small parts of the network, raw FMS carries a strong ceiling
  effect---the null model alone scores 0.82. Pooled across the four families the
  module's level agreement is 0.92, reaching 0.97 on the mildest family
  (Table~\ref{tab:familydetect}).

  \emph{Recall: where quantity-awareness matters
  (Table~\ref{tab:familydetect}).} Reachability is blind to any degradation short of
  disconnection: under \emph{capacity-targeted attack} it misses 46\% of the truly
  critical junctions where the module misses only 10\%, and it misses as many again
  under \emph{source failure}--- reachability recovers just $0.54$
  against the module's $0.92$, even though it is fully precise (always $1.00$) by
  construction. The module holds recall at $0.90$--$0.98$ in every family, paying for
  it in precision (0.73 on average, Table~\ref{tab:familydetect}) via over-warning.

\begin{table}[t]
\centering
\caption{The flow module against two baselines over all eight networks. Every score is a
mean over situations, each counting once, and the \textbf{average} row is the
$n$-weighted mean of the four family rows. \emph{crit.} is the share of scored
junctions that are truly critical. FMS is demand-weighted agreement with WNTR on
\emph{how much} service each junction keeps ($1$ = exact). Reachability scores
precision $1.00$ by construction under the stated connectivity assumptions.}
\label{tab:familydetect}
\setlength{\tabcolsep}{4pt}
\small
\begin{tabular}{@{}lrr@{\hskip 1.1em}ccc@{\hskip 1.1em}cc@{\hskip 1.1em}cc@{}}
\toprule
& & & \multicolumn{3}{c}{Mean FMS}
& \multicolumn{2}{c}{precision}
& \multicolumn{2}{c}{recall}\\
\cmidrule(lr){4-6}
\cmidrule(lr){7-8}
\cmidrule(lr){9-10}
Family & $n$ & crit.
& module & reach. & null
& module & reach.
& module & reach.\\
\midrule
cluster  & 459 & $3.9\%$
& 0.97 & 0.97 & 0.87
& 0.84 & 1.00
& 0.98 & 0.93\\
targeted & 384 & $30.9\%$
& 0.87 & 0.85 & 0.74
& 0.71 & 1.00
& 0.90 & 0.54\\
source   &  24 & $37.2\%$
& 0.78 & 0.73 & 0.65
& 0.51 & 1.00
& 0.92 & 0.54\\
tank     &  72 & $4.2\%$
& 0.87 & 0.98 & 0.95
& 0.20 & 1.00
& 0.94 & 0.92\\
\midrule
\textbf{average} & & $15.8\%$
& \textbf{0.92} & \textbf{0.92} & 0.82
& 0.73 & \textbf{1.00}
& \textbf{0.94} & 0.76\\
\bottomrule
\end{tabular}
\end{table}

\section{Discussion, Limitations and Future Work}\label{sec:discussion}

\emph{Imprecision analysis.} In depth analysis reveals that the imprecisions are generally \emph{over-warning}: recall holds
at $0.90$--$0.98$ across families while precision falls to $0.73$, concentrated in the
two families that force the most rerouting onto backup paths
(Table~\ref{tab:familydetect}). Finer capacity discovery does not fix it
(Supp.~Mat.~S5), and raising capacities outright cures none of the false alarms.

\emph{What the validation does and does not establish.} Section~\ref{sec:validation}
shows that the flow module emulates real hydraulics under the failure distributions we
tested. Beyond physical infrastructure that check is unavailable: no simulator represents
organizational, digital or personnel dependencies, so there is no reference to compare
against, and what stands in its place is inspectability. The logical propagation is locally predictable,
and its attributes let a domain expert encode how the system behaves under stress.

\emph{Limitations.} \emph{No time axis.} Repair and recovery sequencing need a separate, time-stepped
mechanism we do not cover. \emph{No overload cascade.} Capacity is
a hard ceiling: a surviving pipe carries at most its limit, and flow rerouted beyond
that limit goes unserved. The module therefore does not capture cascading damages~\cite{shuang_performance_2014,li_cascading_2025}.
Checking that regime needs a simulator that tracks the network moment by moment through the failure. \emph{Computational Cost.} The module runs slower than a direct WNTR solution (Supp.~Mat.~S8).

\emph{Future work.} The flow module is one plug-in for the proposed
architecture. Power, transport and telecommunications require their own 
domain-specific validation against established simulators which may suggest different approaches.
The logical layer provides the common coupling mechanism, but its
predictions require inspection and comparison with domain experts rather
than physical ground truth. Future work will therefore encode real
organizational and infrastructure systems and assess the resulting
propagation rules with multiple experts.

\emph{Supplementary material.} S1 describes the importer, S2 the
experimental protocol and metric conventions, S3 the required inputs,
S4 the per-network results, S5 the capacity and priority analyses, S6
the ground-truth sensitivity analysis, S7 unsuccessful alternatives,
S8 the computational-cost measurements, and S9 one failure drawn on Net3.

\section{Conclusion}\label{sec:conclusion}
We presented a category-typed propagation architecture in which logical
and quantity-aware dependencies share a discrete functionality scale
and a monotonic update rule. Its water flow module imports EPANET
networks, allocates capacitated supply by fair sharing, and reproduces
PDD-WNTR functionality with a mean Functionality Match Score of 0.92 across eight networks
and four failure families. Compared with reachability, it substantially
improves critical-junction recall for capacity-targeted and source
failures. These results support the use of the architecture as a
light-data approximation of disservice cascades.

\paragraph{Data and Code Availability.} The supplementary material referenced
throughout as Supp.~Mat., together with the code, experimental configuration, and
all data and results, is openly published in the project repository at
\url{https://github.com/cascade-platform-org/CASCADE}; the platform itself runs at
\url{https://app.cascade-platform.org/}.

\bibliographystyle{spmpsci}
\bibliography{complenet-refs}

\end{document}